\documentclass[aps,pr,twocolumn,superscriptaddress,showpacs,preprintnumbers,amsmath,amssymb,fleqn]{revtex4}

\usepackage{graphicx} % Include figure files
\usepackage{dcolumn}  % Align table columns on decimal point

\graphicspath{{ps}}

\begin{document}

%\linenumbers

%\modulolinenumbers[1]

%\pagewiselinenumbers

%\vspace*{-3\baselineskip}
%\resizebox{!}{3cm}{\includegraphics{belle.eps}}

%\preprint{\vbox{ \hbox{   }
%			\hbox{Belle DRAFT {\it 1329-v25}}
%                        \hbox{Intended for {\it PRD}}
%                        \hbox{Author: Y. Kato and T. Iijima}
%                        \hbox{Committee: S. Korper(chair),}
%                        \hbox{B.G. Cheon, J. Yelton}
%  		              % \hbox{hep-ex nnnn}
%}}

\title{ \quad\\[1.0cm] Studies of charmed strange baryons in the $\Lambda D$ final state at Belle}

%%%% >>>>> insert the authorlist here. BEFORE the abstract !!!!! <<<<<
%%%% >>>>> from the authorship confirmation web page
%%% Name the file author.tex and use \input{author} to insert into your latex file.
%\author{Y. Kato and T. Iijima}\affiliation{Kobayashi-Maskawa Institute, Nagoya University, Nagoya 464-8602}
%\input{author}

%%% Paper:    Charmed strange baryons
%%% Journal:  Physical Review D
%%% Contacts: Y. Kato (kato@hepl.phys.nagoya-u.ac.jp)
%%% Non-responding authors or those who said NO are commented out.
%%% ====================================================================
%%% Click the RELOAD button on your web browser to see the updated file.
%%% ====================================================================
%%% Use \input{author} to insert this material into your latex file.
%%%%% Force institutions to appear in alphabetical order when typeset.
\noaffiliation
%%%\affiliation{Aligarh Muslim University, Aligarh 202002}
\affiliation{University of the Basque Country UPV/EHU, 48080 Bilbao}
\affiliation{Beihang University, Beijing 100191}
%%%\affiliation{University of Bonn, 53115 Bonn}
\affiliation{Budker Institute of Nuclear Physics SB RAS, Novosibirsk 630090}
\affiliation{Faculty of Mathematics and Physics, Charles University, 121 16 Prague}
%%%\affiliation{Chiba University, Chiba 263-8522}
\affiliation{Chonnam National University, Kwangju 660-701}
\affiliation{University of Cincinnati, Cincinnati, Ohio 45221}
\affiliation{Deutsches Elektronen--Synchrotron, 22607 Hamburg}
\affiliation{University of Florida, Gainesville, Florida 32611}
%%%\affiliation{Department of Physics, Fu Jen Catholic University, Taipei 24205}
\affiliation{Justus-Liebig-Universit\"at Gie\ss{}en, 35392 Gie\ss{}en}
\affiliation{Gifu University, Gifu 501-1193}
%%%\affiliation{II. Physikalisches Institut, Georg-August-Universit\"at G\"ottingen, 37073 G\"ottingen}
\affiliation{SOKENDAI (The Graduate University for Advanced Studies), Hayama 240-0193}
\affiliation{Gyeongsang National University, Chinju 660-701}
\affiliation{Hanyang University, Seoul 133-791}
\affiliation{University of Hawaii, Honolulu, Hawaii 96822}
\affiliation{High Energy Accelerator Research Organization (KEK), Tsukuba 305-0801}
%%%\affiliation{Hiroshima Institute of Technology, Hiroshima 731-5193}
\affiliation{IKERBASQUE, Basque Foundation for Science, 48013 Bilbao}
%%%\affiliation{University of Illinois at Urbana-Champaign, Urbana, Illinois 61801}
\affiliation{Indian Institute of Science Education and Research Mohali, SAS Nagar, 140306}
\affiliation{Indian Institute of Technology Bhubaneswar, Satya Nagar 751007}
\affiliation{Indian Institute of Technology Guwahati, Assam 781039}
\affiliation{Indian Institute of Technology Madras, Chennai 600036}
\affiliation{Indiana University, Bloomington, Indiana 47408}
\affiliation{Institute of High Energy Physics, Chinese Academy of Sciences, Beijing 100049}
\affiliation{Institute of High Energy Physics, Vienna 1050}
\affiliation{Institute for High Energy Physics, Protvino 142281}
%%%\affiliation{Institute of Mathematical Sciences, Chennai 600113}
\affiliation{INFN - Sezione di Torino, 10125 Torino}
\affiliation{J. Stefan Institute, 1000 Ljubljana}
\affiliation{Kanagawa University, Yokohama 221-8686}
\affiliation{Institut f\"ur Experimentelle Kernphysik, Karlsruher Institut f\"ur Technologie, 76131 Karlsruhe}
%%%\affiliation{Kavli Institute for the Physics and Mathematics of the Universe (WPI), University of Tokyo, Kashiwa 277-8583}
%%%\affiliation{Kennesaw State University, Kennesaw, Georgia 30144}
\affiliation{King Abdulaziz City for Science and Technology, Riyadh 11442}
%%%\affiliation{Department of Physics, Faculty of Science, King Abdulaziz University, Jeddah 21589}
\affiliation{Korea Institute of Science and Technology Information, Daejeon 305-806}
\affiliation{Korea University, Seoul 136-713}
\affiliation{Kyoto University, Kyoto 606-8502}
\affiliation{Kyungpook National University, Daegu 702-701}
\affiliation{\'Ecole Polytechnique F\'ed\'erale de Lausanne (EPFL), Lausanne 1015}
\affiliation{P.N. Lebedev Physical Institute of the Russian Academy of Sciences, Moscow 119991}
\affiliation{Faculty of Mathematics and Physics, University of Ljubljana, 1000 Ljubljana}
\affiliation{Ludwig Maximilians University, 80539 Munich}
%%%\affiliation{Luther College, Decorah, Iowa 52101}
\affiliation{University of Maribor, 2000 Maribor}
\affiliation{Max-Planck-Institut f\"ur Physik, 80805 M\"unchen}
\affiliation{School of Physics, University of Melbourne, Victoria 3010}
%%%\affiliation{Middle East Technical University, 06531 Ankara}
\affiliation{University of Miyazaki, Miyazaki 889-2192}
\affiliation{Moscow Physical Engineering Institute, Moscow 115409}
\affiliation{Moscow Institute of Physics and Technology, Moscow Region 141700}
\affiliation{Graduate School of Science, Nagoya University, Nagoya 464-8602}
\affiliation{Kobayashi-Maskawa Institute, Nagoya University, Nagoya 464-8602}
%%%\affiliation{Nara University of Education, Nara 630-8528}
\affiliation{Nara Women's University, Nara 630-8506}
%%%\affiliation{National Central University, Chung-li 32054}
\affiliation{National United University, Miao Li 36003}
\affiliation{Department of Physics, National Taiwan University, Taipei 10617}
\affiliation{H. Niewodniczanski Institute of Nuclear Physics, Krakow 31-342}
\affiliation{Nippon Dental University, Niigata 951-8580}
\affiliation{Niigata University, Niigata 950-2181}
%%%\affiliation{University of Nova Gorica, 5000 Nova Gorica}
\affiliation{Novosibirsk State University, Novosibirsk 630090}
%%%\affiliation{Osaka City University, Osaka 558-8585}
%%%\affiliation{Osaka University, Osaka 565-0871}
\affiliation{Pacific Northwest National Laboratory, Richland, Washington 99352}
%%%\affiliation{Panjab University, Chandigarh 160014}
%%%\affiliation{Peking University, Beijing 100871}
\affiliation{University of Pittsburgh, Pittsburgh, Pennsylvania 15260}
%%%\affiliation{Punjab Agricultural University, Ludhiana 141004}
%%%\affiliation{Research Center for Electron Photon Science, Tohoku University, Sendai 980-8578}
%%%\affiliation{Research Center for Nuclear Physics, Osaka University, Osaka 567-0047}
%%%\affiliation{RIKEN BNL Research Center, Upton, New York 11973}
%%%\affiliation{Saga University, Saga 840-8502}
\affiliation{University of Science and Technology of China, Hefei 230026}
\affiliation{Seoul National University, Seoul 151-742}
%%%\affiliation{Shinshu University, Nagano 390-8621}
\affiliation{Showa Pharmaceutical University, Tokyo 194-8543}
\affiliation{Soongsil University, Seoul 156-743}
%%%\affiliation{University of South Carolina, Columbia, South Carolina 29208}
\affiliation{Sungkyunkwan University, Suwon 440-746}
\affiliation{School of Physics, University of Sydney, New South Wales 2006}
\affiliation{Department of Physics, Faculty of Science, University of Tabuk, Tabuk 71451}
\affiliation{Tata Institute of Fundamental Research, Mumbai 400005}
\affiliation{Excellence Cluster Universe, Technische Universit\"at M\"unchen, 85748 Garching}
\affiliation{Department of Physics, Technische Universit\"at M\"unchen, 85748 Garching}
\affiliation{Toho University, Funabashi 274-8510}
%%%\affiliation{Tohoku Gakuin University, Tagajo 985-8537}
\affiliation{Department of Physics, Tohoku University, Sendai 980-8578}
\affiliation{Earthquake Research Institute, University of Tokyo, Tokyo 113-0032}
\affiliation{Department of Physics, University of Tokyo, Tokyo 113-0033}
\affiliation{Tokyo Institute of Technology, Tokyo 152-8550}
\affiliation{Tokyo Metropolitan University, Tokyo 192-0397}
%%%\affiliation{Tokyo University of Agriculture and Technology, Tokyo 184-8588}
%%%\affiliation{University of Torino, 10124 Torino}
%%%\affiliation{Toyama National College of Maritime Technology, Toyama 933-0293}
\affiliation{Utkal University, Bhubaneswar 751004}
\affiliation{Virginia Polytechnic Institute and State University, Blacksburg, Virginia 24061}
\affiliation{Wayne State University, Detroit, Michigan 48202}
\affiliation{Yamagata University, Yamagata 990-8560}
\affiliation{Yonsei University, Seoul 120-749}
  \author{Y.~Kato}\affiliation{Kobayashi-Maskawa Institute, Nagoya University, Nagoya 464-8602} % Nagoya
  \author{T.~Iijima}\affiliation{Kobayashi-Maskawa Institute, Nagoya University, Nagoya 464-8602}\affiliation{Graduate School of Science, Nagoya University, Nagoya 464-8602} % Nagoya
% \author{A.~Abdesselam}\affiliation{Department of Physics, Faculty of Science, University of Tabuk, Tabuk 71451} % Tabuk
  \author{I.~Adachi}\affiliation{High Energy Accelerator Research Organization (KEK), Tsukuba 305-0801}\affiliation{SOKENDAI (The Graduate University for Advanced Studies), Hayama 240-0193} % KEK
% \author{K.~Adamczyk}\affiliation{H. Niewodniczanski Institute of Nuclear Physics, Krakow 31-342} % Krakow
  \author{H.~Aihara}\affiliation{Department of Physics, University of Tokyo, Tokyo 113-0033} % Tokyo
% \author{S.~Al~Said}\affiliation{Department of Physics, Faculty of Science, University of Tabuk, Tabuk 71451}\affiliation{Department of Physics, Faculty of Science, King Abdulaziz University, Jeddah 21589} % Tabuk
% \author{K.~Arinstein}\affiliation{Budker Institute of Nuclear Physics SB RAS, Novosibirsk 630090}\affiliation{Novosibirsk State University, Novosibirsk 630090} % BINP
% \author{Y.~Arita}\affiliation{Graduate School of Science, Nagoya University, Nagoya 464-8602} % Nagoya
  \author{D.~M.~Asner}\affiliation{Pacific Northwest National Laboratory, Richland, Washington 99352} % PNNL
% \author{T.~Aso}\affiliation{Toyama National College of Maritime Technology, Toyama 933-0293} % Toyama
% \author{H.~Atmacan}\affiliation{Middle East Technical University, 06531 Ankara} % METU
  \author{V.~Aulchenko}\affiliation{Budker Institute of Nuclear Physics SB RAS, Novosibirsk 630090}\affiliation{Novosibirsk State University, Novosibirsk 630090} % BINP
% \author{T.~Aushev}\affiliation{Moscow Institute of Physics and Technology, Moscow Region 141700} % MIPT
  \author{R.~Ayad}\affiliation{Department of Physics, Faculty of Science, University of Tabuk, Tabuk 71451} % Tabuk
% \author{T.~Aziz}\affiliation{Tata Institute of Fundamental Research, Mumbai 400005} % Tata
% \author{V.~Babu}\affiliation{Tata Institute of Fundamental Research, Mumbai 400005} % Tata
  \author{I.~Badhrees}\affiliation{Department of Physics, Faculty of Science, University of Tabuk, Tabuk 71451}\affiliation{King Abdulaziz City for Science and Technology, Riyadh 11442} % Tabuk
% \author{S.~Bahinipati}\affiliation{Indian Institute of Technology Bhubaneswar, Satya Nagar 751007} % IITB
  \author{A.~M.~Bakich}\affiliation{School of Physics, University of Sydney, New South Wales 2006} % Sydney
% \author{A.~Bala}\affiliation{Panjab University, Chandigarh 160014} % Panjab
% \author{Y.~Ban}\affiliation{Peking University, Beijing 100871} % Peking
% \author{V.~Bansal}\affiliation{Pacific Northwest National Laboratory, Richland, Washington 99352} % PNNL
  \author{E.~Barberio}\affiliation{School of Physics, University of Melbourne, Victoria 3010} % Melbourne
% \author{M.~Barrett}\affiliation{University of Hawaii, Honolulu, Hawaii 96822} % Hawaii
% \author{W.~Bartel}\affiliation{Deutsches Elektronen--Synchrotron, 22607 Hamburg} % DESY
% \author{A.~Bay}\affiliation{\'Ecole Polytechnique F\'ed\'erale de Lausanne (EPFL), Lausanne 1015} % Lausanne
% \author{I.~Bedny}\affiliation{Budker Institute of Nuclear Physics SB RAS, Novosibirsk 630090}\affiliation{Novosibirsk State University, Novosibirsk 630090} % BINP
% ...
% Ljubljana
% \author{N.~Zwahlen}\affiliation{\'Ecole Polytechnique F\'ed\'erale de Lausanne (EPFL), Lausanne 1015} % Lausanne
% \author{O.~Zyukova}\affiliation{Budker Institute of Nuclear Physics SB RAS, Novosibirsk 630090}\affiliation{Novosibirsk State University, Novosibirsk 630090} % BINP
\collaboration{The Belle Collaboration}

%\collaboration{The Belle Collaboration}
%\noaffiliation
%% end author list

\begin{abstract}
We report the discovery of $\Xi_{c}(3055)^{0}$, observed by its decay into the final state $\Lambda D^{0}$, and 
present the first observation and evidence of the decays of $\Xi_c(3055)^{+}$ and $\Xi_c(3080)^{+}$ into $\Lambda D^{+}$.
We also perform a combined analysis of the $\Lambda D^{+}$ with the $\Sigma_{c}^{++}K^{-}$ and $\Sigma_{c}^{\ast ++}K^{-}$ decay modes
to measure the ratios of branching fractions, masses and widths with improved accuracy.
We measure the ratios of branching fractions ${\cal B}(\Xi_{c}(3055)^{+} \to \Lambda D^{+})/{\cal B}(\Xi_{c}(3055)^{+} \to \Sigma_{c}^{++}K^{-})=5.09\pm1.01\pm0.76$,
${\cal B}(\Xi_{c}(3080)^{+} \to \Lambda D^{+})/{\cal B}(\Xi_{c}(3080)^{+} \to \Sigma_{c}^{++}K^{-})=1.29\pm0.30\pm0.15$, and 
${\cal B}(\Xi_{c}(3080)^{+} \to \Sigma_{c}^{\ast ++}K^{-} )/{\cal B}(\Xi_{c}(3080)^{+} \to \Sigma_{c}^{++}K^{-})=1.07\pm0.27\pm0.01$,
where the uncertainties are statistical and systematic.
The analysis is performed using a 980 fb$^{-1}$ data sample collected with the
Belle detector at the KEKB asymmetric-energy $e^{+}e^{-}$ collider.

\end{abstract}

\pacs{13.30.-a, 14.20.-c}

\maketitle

%%%% >>>> keep the final version single-spaced
\tighten

{\renewcommand{\thefootnote}{\fnsymbol{footnote}}}
\setcounter{footnote}{0}

\section{Introduction}\label{section_intro}
 In recent years, there has been significant progress in the study of the 
charmed baryon spectrum, mainly from the Belle and BaBar experiments \citep{Mizuk:2004yu,Chistov:2006zj,
Aubert:2006je,Aubert:2006sp,Abe:2006rz,Aubert:2007dt,Kato:2013ynr,Solovieva:2008fw,Lesiak:2008wz}.
In the charmed strange baryon sector, a number of excited states ($\Xi_c^{\ast}$) have been observed.
Belle reported evidence for two
excited states, $\Xi_c(2980)$ and $\Xi_c(3080)$, in the $\Lambda_{c}^{+}K^{-}\pi^{+}$
and $\Lambda_{c}^{+}K_{S}^{0}\pi^{-}$ final states \cite{Chistov:2006zj}.
These states have been confirmed by BaBar \cite{Aubert:2007dt}. In the same paper, 
BaBar also claimed evidence for two resonances, the $\Xi_c(3055)^{+}$ and the $\Xi_c(3123)^{+}$, observed 
in the $\Sigma_{c}^{++}K^{-}$ and $\Sigma_{c}^{\ast ++}K^{-}$
final states. Recently, Belle confirmed the existence of the $\Xi_c(3055)^{+}$,
but no evidence was found for the $\Xi_c(3123)^{+}$ \cite{Kato:2013ynr}.
As discussed in Refs.~\cite{Cheng:2006dk,Liu:2012sj}, the decay pattern of 
charmed baryons provides an important contribution to our understanding of the nature of the states.
To date, all measurements of $\Xi_{c}^{\ast}$ baryons were performed using decays in which 
the charm quark is contained in the final-state baryon. Measurements of final states 
in which the charm quark is part of the final state meson provide complementary information. 

In this paper, we report studies of $\Xi_{c}^{\ast}$ baryons decaying to the $\Lambda D^{+}$ and $\Lambda D^{0}$
final states using a data sample with an integrated luminosity of 980 fb$^{-1}$ collected with the
Belle detector at the KEKB asymmetric-energy $e^{+}e^{-}$ collider. We find significant signals for
$\Xi_{c}(3055)^{+}$ and $\Xi_{c}(3080)^{+}$ decays into $\Lambda D^{+}$.
In the $\Lambda D^{0}$ final state, we report observation of the $\Xi_{c}(3055)^{0}$.
These measurements constitute the first observation and evidence for the $\Xi_{c}(3055)$
and $\Xi_{c}(3080)$ into the $\Lambda D$ final states, and the first-ever observation of the $\Xi_{c}(3055)^{0}$.
We also perform a combined analysis of the $\Lambda D^{+}$ , $\Sigma_{c}^{++}K^{-}$, and $\Sigma_{c}^{\ast ++}K^{-}$ final states to measure the ratios of branching fractions and to
improve the accuracy of the mass and width measurements.
%We find the mass of the $\Xi_{c}(3080)^{+}$ in the $\Lambda D^{+}$ decay mode is found to be inconsistent with 
%$\Sigma_{c}^{++}K^{-}$, and $\Sigma_{c}^{\ast ++}K^{-}$ final states by about 2 MeV/$c^{2}$.
%We determine the combined masses of the $\Xi_{c}(3055)^{+}$ and $\Xi_{c}(3080)^{+}$ by just taking the average and 
%scale the uncertainty by $\sqrt(\chi^{2}/ndf$, where $ndf$ is the number of degree of freedom.
%We determine the ratios of branching fractions and combined widths by simultaneous fit between three decay modes with common width.

The remaining sections of the paper are organized as follows. 
In Sections \ref{section_data} and \ref{section_selection}, we describe the data sample and event selections.
In Section \ref{section_signal}, observations and measurements of $\Xi_{c}^{\ast}$ baryons
in the $\Lambda D^{+}$ and $\Lambda D^{0}$ final states are presented.
In Section \ref{section_combine}, the combined analysis with the $\Sigma_{c}^{++}K^{-}$ and $\Sigma_{c}^{\ast ++}K^{-}$ final states
is presented. Finally, the summary and conclusion are given.

\section{Data samples and the Belle detector}\label{section_data}
%We use a data sample with a total integrated luminosity of 980 fb$^{-1}$ recorded with 
%the Belle detector at the KEKB asymmetric beam energy $e^{+}e^{-}$ collider \cite{KEKB}. Data samples with 
%different beam energies are combined in this study: at the $\Upsilon(4S)$ ($\sqrt{s}$ = 10.58 GeV, 711 fb$^{-1}$) and 
%near it (89.5 fb$^{-1}$), at the $\Upsilon(5S)$ ($\sqrt{s}$ = 10.88 GeV, 121.0 fb$^{-1}$) and near it (29.3 fb$^{-1}$),
%at the $\Upsilon(2S)$ ($\sqrt{s}$ = 10.02 GeV, 24.9 fb$^{-1}$) and near it (1.8 fb$^{-1}$),
%at the $\Upsilon(1S)$ ($\sqrt{s}$ = 9.46 GeV, 5.7 fb$^{-1}$) and near it (1.8 fb$^{-1}$), and 
%at the $\Upsilon(3S)$ ($\sqrt{s}$ = 10.36 GeV, 2.9 fb$^{-1}$) and near it (0.3 fb$^{-1}$).\par
We use a data sample with a total integrated luminosity of 980 fb$^{-1}$ recorded with 
the Belle detector at the KEKB asymmetric-beam-energy $e^{+}e^{-}$ collider \cite{KEKB}.
The Belle detector is a large-solid-angle magnetic
spectrometer that consists of a silicon vertex detector (SVD),
a 50-layer central drift chamber (CDC), an array of
aerogel threshold Cherenkov counters (ACC),  % <- \v{C}erenkov 2007.08
a barrel-like arrangement of time-of-flight
scintillation counters (TOF), and an electromagnetic calorimeter
comprised of CsI(Tl) crystals (ECL) located inside 
a superconducting solenoid coil that provides a 1.5~T
magnetic field.  An iron flux-return located outside of
the coil is instrumented to detect $K_L^0$ mesons and to identify
muons.  The detector is described in detail elsewhere~\cite{Belle}.
Two inner detector configurations were used. A 2.0-cm radius beampipe
and a 3-layer silicon vertex detector was used for the first sample
of 156 fb$^{-1}$, while a 1.5-cm radius beampipe, a 4-layer
silicon detector and a small-cell inner drift chamber were used to record
the remaining 824 fb$^{-1}$ \cite{svd2}.\par

We use a GEANT-based Monte Carlo (MC) simulation \cite{GEANT} to model the detector response
and its acceptance to obtain the reconstruction efficiency and the mass resolution for the signal.
We re-weight the signal MC sample according to the scaled-momentum $x_{p}=p^{\ast}/p_{\rm max}$ distributions,
based on the measurements in real data, to obtain the correct reconstruction efficiency.
Here, $p^{\ast}$ is the momentum of the $\Xi_{c}^{\ast}$ system
in the center-of-mass frame and $p_{\rm max} = \sqrt{s/4-M^{2}c^{4}}/c$, where $s$ is the total center-of-mass 
energy squared, $M$ is the invariant mass of the $\Xi_{c}^{\ast}$ system, and $c$ is the speed of light.
We also use MC events generated with EVTGEN \cite{evtgen} and JETSET \cite{jetset}
to study the mass distribution in the background process $e^{+}e^{-} \to q\bar{q}$ process ($q=u,d,s,c$ and $b$).

\section{Event selection}\label{section_selection}
Our analysis is optimized to search for decays of $\Xi_{c}^{\ast}$ baryons into the $\Lambda D^{+}$ and $\Lambda D^{0}$ final states.
Throughout this paper, the inclusion of the charge-conjugate decay mode is implied
unless otherwise stated.
A $\Lambda$ candidate is reconstructed via its decay into $p \pi^{-}$.
A $D^{+}$ candidate is reconstructed via its decay into $K^{-}\pi^{+}\pi^{+}$.
A $D^{0}$ candidate is reconstructed via its decay into $K^{-}\pi^{+}$, $K^{-}\pi^{+}\pi^{+}\pi^{-}$, and $K^{-}\pi^{+}\pi^{0}$.

The selection of charged hadrons is based on information from the tracking system (SVD and CDC) and 
hadron identification system (CDC, ACC, and TOF). The charged hadrons that are not 
associated with the $\Lambda$ candidate are required to have a point of closest approach to the interaction point
that is within 2 cm along the $z$ axis and within 0.2 cm in the transverse ($r$-$\phi$) plane. 
The $z$ axis is opposite the positron beam direction. For each track, likelihood values $\mathcal{L}_{p}$, $\mathcal{L}_{K}$, and $\mathcal{L}_{\pi}$
are provided by the hadron identification system, based on the ionization losses in the CDC,
the number of detected Cherenkov photons in the ACC, and the time-of-flight measured by the TOF.
The likelihood ratio is defined as $\mathcal{L}(i:j)=\mathcal{L}_{i}/(\mathcal{L}_{i}+\mathcal{L}_{j})$.
A track is identified as a proton if the likelihood ratios $\mathcal{L}(p:\pi)$ and $\mathcal{L}(p:K)$ are greater than 0.6,
as a kaon if the likelihood ratios $\mathcal{L}(K:\pi)$ and $\mathcal{L}(K:p)$ are greater than 0.6, or as a
pion if the likelihood ratios $\mathcal{L}(\pi:K)$ and $\mathcal{L}(\pi:p)$ are greater than 0.6.
In addition, an electron likelihood is provided based on information from the ECL, ACC, and CDC \cite{eid}.
A track with an electron likelihood greater than 0.95 is rejected. \par
The momentum-averaged efficiencies of hadron identification are about 90$\%$, 90$\%$, and 93$\%$ for pions,
kaons, and protons, respectively.
%The momentum averaged probability to misidentify a pion (kaon) track as a kaon (pion) track is about 9 (10)$\%$, and the momentum averaged probability to
%misidentify a pion or kaon track as a proton track is about 5$\%$.
The momentum-averaged probability to misidentify a pion as a kaon is about 9$\%$,
to misidentify a kaon as a pion about 10$\%$, and to 
misidentify a pion or kaon as a proton about 5$\%$.
The $\pi^{0}$ candidates are reconstructed from pairs of photons whose invariant mass ($M_{\gamma \gamma}$) satisfies
$120$ MeV/$c^{2}<M_{\gamma \gamma}<150$ MeV/${\it c}$$^{2}$, which corresponds to $\pm 2.5 \sigma$ (where $\sigma$ is the one-standard-deviation of the resolution).
The energy of each photon in the laboratory frame is required to be greater than $50$ MeV
and the energy of the $\pi^{0}$ candidate in the laboratory frame is required to be greater than 500 MeV.
The $D^{+}$ candidates are selected by requiring $|M(K^{-}\pi^{+}\pi^{+})-m_{D^{+}}|<$ 12 MeV/${\it c}$$^{2}$, 
where $m_{D^{+}}$ is the $D^{+}$ mass \cite{PDG}.
The $D^{0}$ candidates for each decay mode of the $D^{0}$ are selected by requiring 
$|M(K^{-}\pi^{+})-m_{D^{0}}|<$ 14 MeV/${\it c}$$^{2}$,
$|M(K^{-}\pi^{+}\pi^{+}\pi^{-})-m_{D^{0}}|<$ 11 MeV/${\it c}$$^{2}$, 
and $|M(K^{-}\pi^{+}\pi^{0})-m_{D^{0}}|<$ 27 MeV/${\it c}$$^{2}$, where $m_{D^{0}}$ is the $D^{0}$ mass.
These mass ranges correspond to $\pm 2.5\sigma$.
To improve the momentum resolution, the daughter particles are fitted to a
common vertex together with an invariant mass constrained to the $D^{+}$ or $D^{0}$ mass.
%The $\Lambda$ candidates are selected based on their decay vertex information \cite{Lambdaselection}
The $\Lambda$ candidates are selected using cuts on four parameters:
the angular difference between the $\Lambda$ flight direction and
the direction pointing from IP to the decay vertex in the transverse plane; the distance between each track and
the IP in the transverse plane; the distance between the decay vertex and the IP in the transverse plane; and
the displacement along $z$ of the closest-approach points of the two tracks to the beam axis. 
Also, the invariant mass of a $\Lambda$ candidate is required to be within 3 MeV/${\it c}$$^{2}$ of the 
$\Lambda$ mass, which corresponds to $\pm 3 \sigma$.
Excited charmed baryons are known to be produced with much higher average momenta than the 
combinatorial background. We thus require that $x_{p}$ be greater than 
0.7 for the $\Lambda D^{+}$ and 0.8 for the $\Lambda D^{0}$ modes.

\section{Observation of $\Xi_{c}t^{\ast}\to\Lambda D$ decays}\label{section_signal}
\begin{figure*}[htbp]
  \begin{center}
    \includegraphics[scale=0.3]{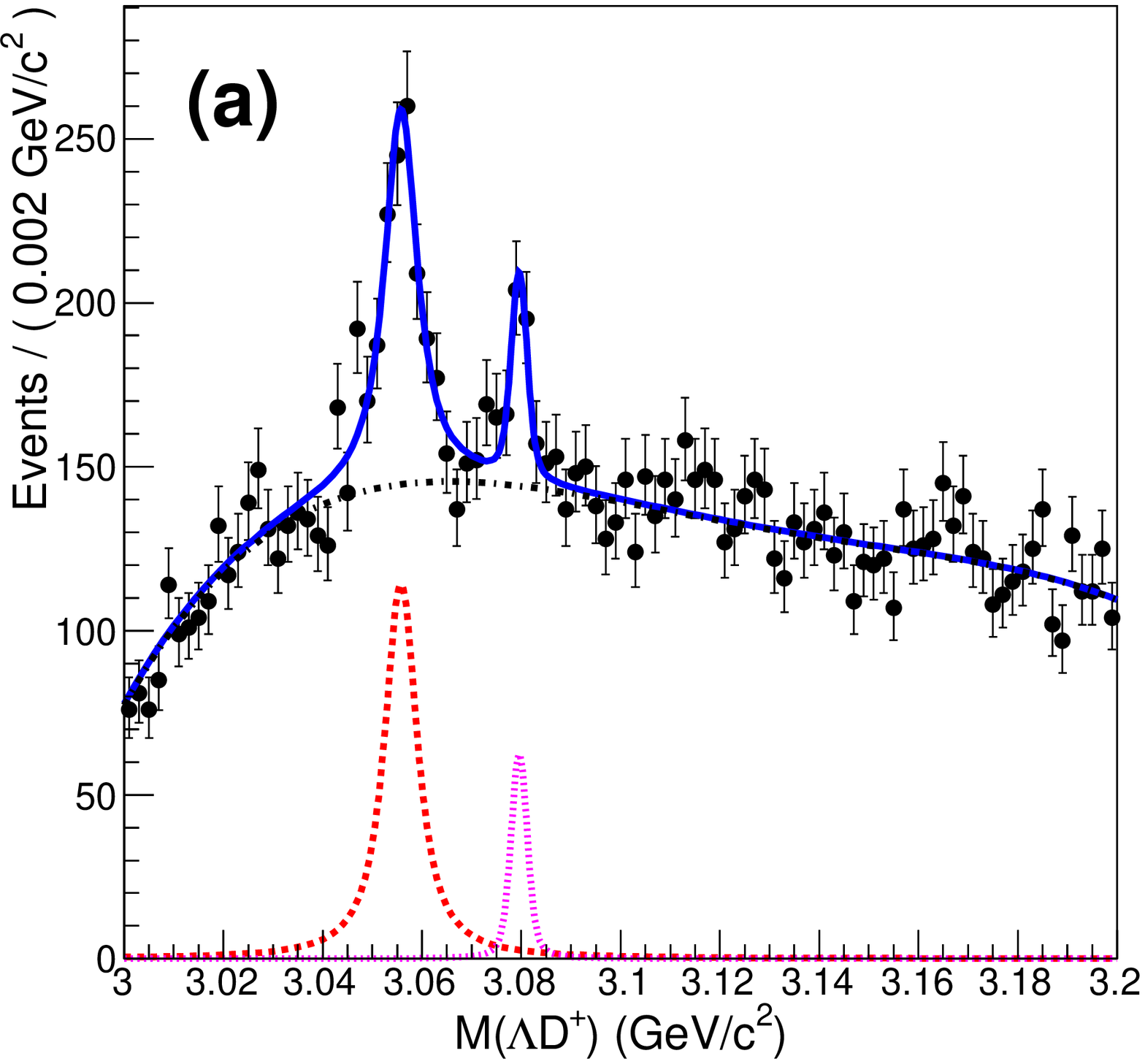}
    \includegraphics[scale=0.3]{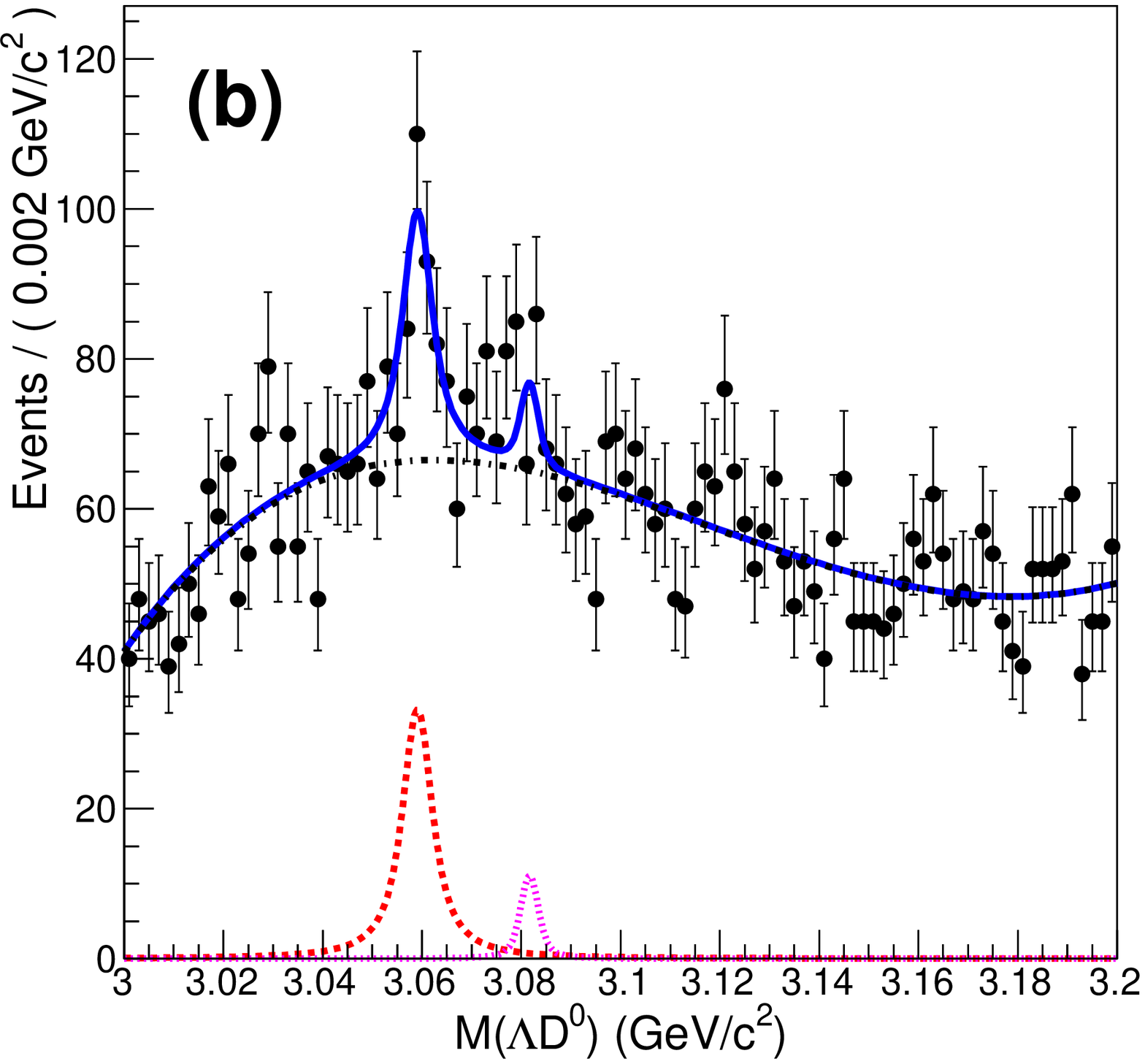}
    \includegraphics[scale=0.3]{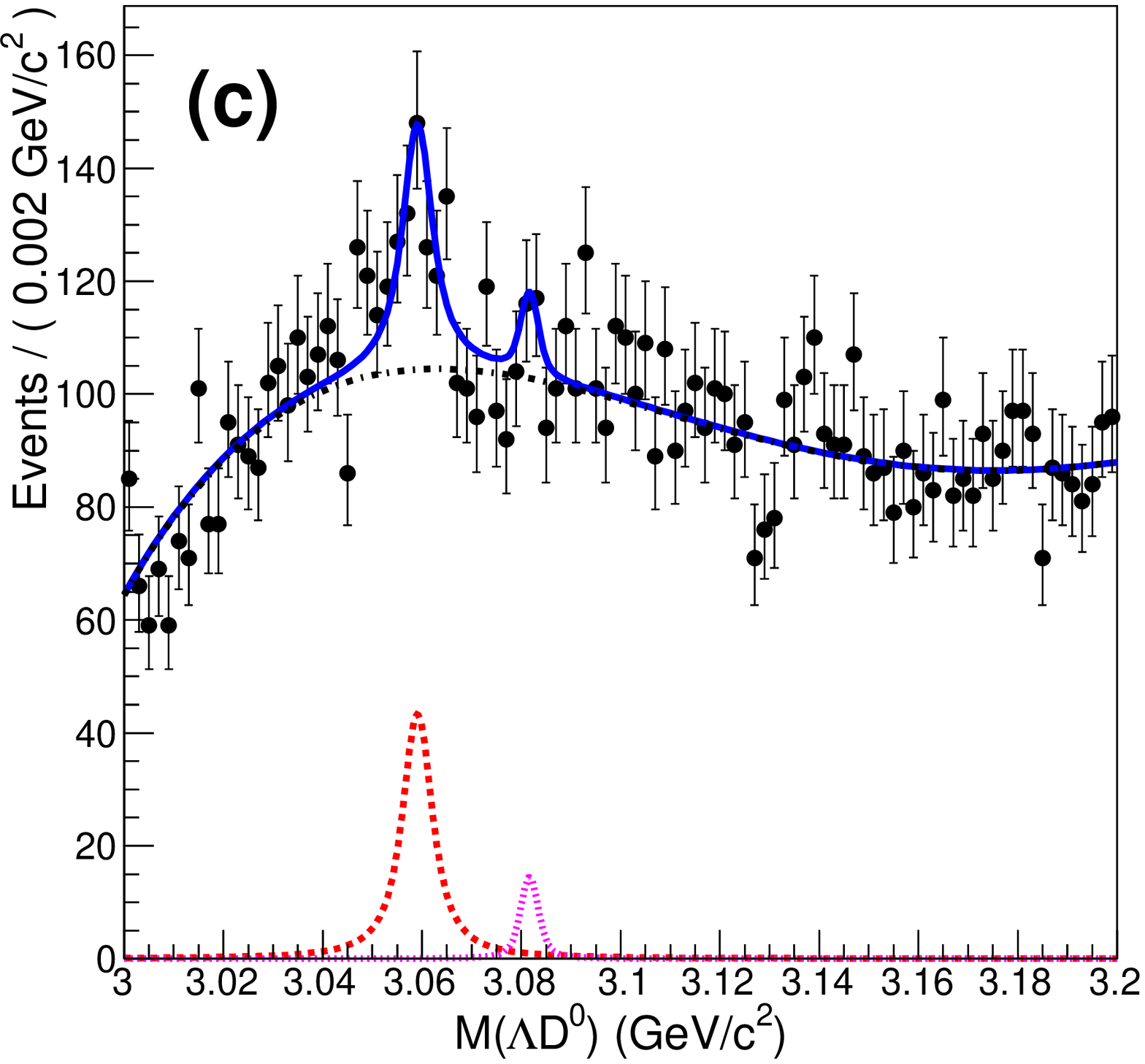}
    \includegraphics[scale=0.3]{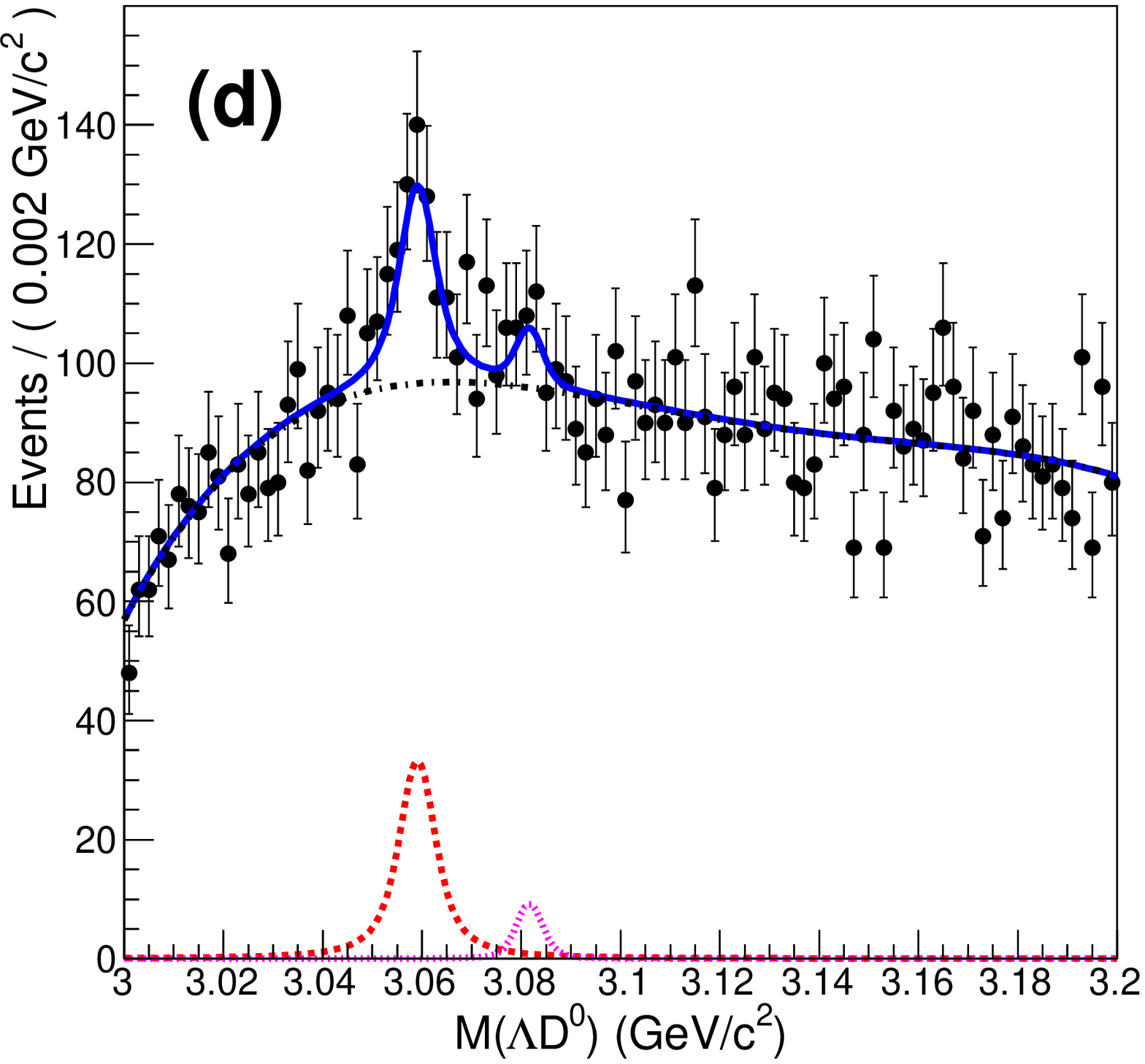}
    \caption{$M(\Lambda D)$ distributions. Points with statistical error bars are data.
             Blue solid lines show the fit results. The red dashed, magenta dotted, and black dashed-dotted lines show
             the $\Xi_{c}(3055)$ signal, the $\Xi_{c}(3080)$ signal, and the background components, respectively.
             (a) $M(\Lambda D^{+})$ distribution; $M(\Lambda D^{0})$ distributions for the (b) $K^{-}\pi^{+}$,
             (c) $K^{-}\pi^{+}\pi^{+}\pi^{-}$, and (d) $K^{-}\pi^{+}\pi^{0}$ $D^{0}$ decay modes.}
    \label{mlamd}
  \end{center}
\end{figure*}

Figure \ref{mlamd} shows the $\Lambda D$ invariant-mass ($M(\Lambda D)$) distributions
for data after the application of all the selection criteria; signals near 3055
and 3080 MeV/${\it c}$$^{2}$ are seen. We do not observe any such peaks in the 
distributions in wrong-sign $\overline{\Lambda} D$ combinations, in data from the $D$ meson mass sideband, nor in MC events that do not include these resonances.
Hereinafter, $\Xi_{c}^{\ast}$ baryons corresponding to these peaks are referred to as $\Xi_c(3055)$ and $\Xi_{c}(3080)$. 
In order to evaluate the masses, widths, and statistical significances of the $\Xi_{c}^{\ast}$ states, 
we apply an unbinned extended maximum likelihood (UML) fit to the mass spectra in the invariant mass range of 3.0$-$3.2 GeV/$c^{2}$.
For the $\Lambda D^{0}$ mode, the fit is performed simultaneously for the three different $D^{0}$ decay modes,
with their relative yields fixed using the product of their known branching fractions \cite{PDG} and detection efficiencies.
The masses and widths of the $\Xi_{c}^{\ast}$ states are constrained to be the same for all modes.
The detection efficiencies for the $\Xi_{c}(3055)^{0}$ and $\Xi_{c}(3080)^{0}$ are found to exhibit no difference within the statistical
precision of the MC sample, which is smaller than $1\%$. Therefore, we use common efficiency values for these states.
The relative yields are fixed to $K^{-}\pi^{+}$:$K^{-}\pi^{+}\pi^{+}\pi^{-}$:$K^{-}\pi^{+}\pi^{0}$= 1.00:1.30:1.15.
The probability density functions (PDF) for the $\Xi_{c}^{\ast}$ components are represented by convolutions of Breit-Wigner shapes 
with Gaussian distributions to take the intrinsic invariant mass resolution, $\sigma_{\rm res}$, into account.
Using the signal MC events, we determine $\sigma_{\rm res}$ for the $\Lambda D^{+}$ mode to be
1.1 MeV/${\it c}$$^{2}$ for the $\Xi_c(3055)^{+}$ and 1.3 MeV/${\it c}$$^{2}$ for the $\Xi_c(3080)^{+}$.
%In the $\Lambda D^{0}$ mode of the $D^{0}$ decay modes of $K^{-}\pi^{+}$, $K^{-} \pi^{+} \pi^{+} \pi^{-}$, and $K^{-} \pi^{+} \pi^{0}$, we
%determine the $\sigma_{\rm res}$ to be 1.1 MeV/${\it c}$$^{2}$, 1.1 MeV/${\it c}$$^{2}$, and  2.0 MeV/${\it c}$$^{2}$, respectively, 
%for the $\Xi_c(3055)^{0}$ and 1.3 MeV/${\it c}$$^{2}$, 1.3 MeV/${\it c}$$^{2}$, and  2.2 MeV/${\it c}$$^{2}$, respectively,
%for the $\Xi_c(3080)^{0}$.
In the $\Lambda D^{0}$ mode, we determine $\sigma_{\rm res}$ to be 1.1 and 2.0 MeV/${\it c}$$^{2}$
for the $D^{0}$ decay mode without and with $\pi^{0}$, respectively for the $\Xi_{c}(3055)^{0}$
and 1.3 and 2.2 MeV/${\it c}$$^{2}$ for the $\Xi_{c}(3080)^{0}$.
The masses, widths and yields of the $\Xi_{c}^{\ast}$ states are treated as free parameters.
A third-order Chebyshev polynomial is used to model the PDF for the combinatorial background.
The statistical significance is evaluated from $-2\ln{(\mathcal{L}_{0}/\mathcal{L})}$,
where $\mathcal{L}_{0}$ ($\mathcal{L}$) is the likelihood for the fit without (with) the signal component. When we evaluate $\mathcal{L}_{0}$ for one of the $\Xi_c^{\ast}$ states, the other $\Xi_c^{\ast}$
state is included in the fit. The $-2\ln{(\mathcal{L}_{0}/\mathcal{L})}$ values are 144.6 for the $\Xi_c(3055)^{+}$, 30.0 for the $\Xi_c(3080)^{+}$,
83.1 for the $\Xi_c(3055)^{0}$, and 6.6 for the $\Xi_c(3080)^{0}$.
By taking into account the change by 3 of the number of degrees of freedom in the UML fit
associated with the inclusion of the $\Xi_c^{\ast}$ states, the statistical significances are
11.7$\sigma$, 4.8$\sigma$, 8.6$\sigma$, and 1.7$\sigma$
for the $\Xi_c(3055)^{+}$, $\Xi_c(3080)^{+}$, $\Xi_c(3055)^{0}$, and $\Xi_c(3080)^{0}$, respectively.
The peak for the $\Xi_c(3080)^{0}$ is not statistically significant.

We estimate the systematic uncertainty of the masses and widths of $\Xi_{c}(3055)^{0}$, $\Xi_{c}(3055)^{+}$ and $\Xi_{c}(3080)^{+}$ 
in the $\Lambda D^{+}$ decay mode as the changes produced by giving reasonable variations to the fitting technique.
The stability of the background shape is checked by changing the fit region and background PDF.
The maximum deviation from the nominal fit is taken as the systematic uncertainty.
To check the uncertainty due to $\sigma_{\rm res}$, the ratio $r_{\sigma}=\sigma_{D}^{\rm MC}/ \sigma_{D}^{\rm data}$
is evaluated, where $\sigma_{D}^{\rm MC}$ and $\sigma_{D}^{\rm data}$ are the $D^{0}$ mass resolution for MC and data.
For the $\Lambda D^{0}$ mode, $r_{\sigma}$ is 1.16, 1.16 and 1.08 for the $D^{0}$ final state of 
$K^{-}\pi^{+}$, $K^{-}\pi^{+}\pi^{+}\pi^{-}$ and $K^{-}\pi^{+}\pi^{0}$, respectively.
We evaluate the uncertainty by fitting data with $\sigma_{\rm res}$ scaled by 16$\%$ for all the decay modes.
To check the uncertainty on the mass due to a possible mis-calibration of the momentum and energy measurements, 
we check the reconstructed $D^{0}$ masses for both data and signal MC.
In each mode, the peak position is observed to have a distinct but small deviation from the world average.
However, these deviations are well reproduced by the MC and, because of the mass-constrained fit,
have little effect on the determination of the masses of the $\Xi_c^{\ast}$ baryons. In the signal MC, the 
differences between the input and output masses of the $\Xi_{c}^{\ast}$ baryons is less than 0.1 MeV/$c^{2}$
for all $D^{0}$ decay modes. We assign a systematic uncertainty of 0.1 MeV$/c^{2}$
on the mass measurements. We perform fits that include the interference of the $\Xi_{c}(3055)$ and $\Xi_{c}(3080)$
by introducing the phase between two Breit-Wigner amplitudes.
The systematic uncertainties are summarized in Table \ref{summary_sys_lamd0}.
The fit result for the $\Xi_{c}(3080)^{+}$ width is ($1.4\pm1.8$) MeV, which is consistent with zero.
Therefore, we set a 90$\%$ confidence level upper limit on the width.
We redo the fit by changing the width; the width for which
the likelihood ratio $-2\ln{(\mathcal{L}/\mathcal{L}_{0 \Gamma})}$ is 2.7, where $L_{0 \Gamma}$
is the likelihood with the zero width for $\Xi_{c}(3080)^{+}$, is assigned as the 90$\%$ 
confidence level upper limit. We obtain the upper limit $\Gamma_{\Xi_{c}(3080)^{+}}<6.3$ MeV.
The measurements of the masses and widths are summarized in Table \ref{summary_mass_lambdad}.
Note that the final values for the $\Xi_{c}(3055)^{+}$ and $\Xi_{c}(3080)^{+}$ masses and widths
in this paper are those combined with $\Sigma_{c}^{++}K^{-}$ and $\Sigma_{c}^{\ast ++}K^{-}$ decay modes.
Values in the $\Lambda D$ mode only are shown to compare with other decay modes.
We find that the mass of the $\Xi_{c}(3055)^{+}$ and widths of $\Xi_{c}(3055)^{+}$ and $\Xi_{c}(3080)^{+}$
are consistent with our previous measurements with the $\Sigma_{c}^{++}K^{-}$ and $\Sigma_{c}^{\ast ++}K^{-}$ decay modes
\cite{Kato:2013ynr}. However, we find a small inconsistency for the mass of the $\Xi_{c}(3080)^{+}$, which may indicate the 
possible underestimation of the systematic uncertainty for the determination of the masses.
We determine the combined value for the masses of the $\Xi_{c}(3055)^{+}$ and $\Xi_{c}(3080)^{+}$ by taking the weighted average.
The uncertainty is scaled by $\sqrt{\chi^{2}/(N-1)}$, where $N$ is the number of different decay modes, which is 2 for $\Xi_{c}(3055)^{+}$ and 3 for $\Xi_{c}(3080)^{+}$,
if the $\chi^{2}/(N-1)$ is greater than one; this is the recipe used in Ref.~\cite{PDG}.
The scale factor for the $\Xi_c(3055)^{+}$ is 1.0 and that for the $\Xi_c(3080)^{+}$ is 3.3.
The measured mass of the $\Xi_{c}(3055)^{+}$ is $(3055.9\pm0.4)$ MeV/$c^{2}$ and that for $\Xi_{c}(3080)^{+}$ is $(3077.9\pm0.9)$ MeV/$c^{2}$. The combined values for the widths are determined by simultaneous fit
with $\Sigma_{c}^{++}K^{-}$ and $\Sigma_{c}^{\ast ++}K^{-}$ decay modes as described in the next section.

\begin{center}
  \begin{table*}[htbp]
    \caption{Systematic uncertainties for the mass (MeV/$c^{2}$) and width (MeV) of the $\Xi_{c}^{\ast}$.}
     \begin{tabular}{c|c@{\hspace{0.2cm}}c@{\hspace{0.2cm}}c@{\hspace{0.2cm}}c@{\hspace{0.2cm}}c@{\hspace{0.2cm}}c@{\hspace{0.2cm}}c@{\hspace{0.2cm}}c} \hline \hline
       Source & M$_{\Xi_c(3055)^{0}}$  & $\Gamma_{\Xi_c(3055)^{0}}$ & M$_{\Xi_c(3055)^{+}}$  & $\Gamma_{\Xi_c(3055)^{+}}$ & M$_{\Xi_c(3080)^{+}}$\\ \hline
       Background shape    & 0.6                  & 1.0                      &   0.1         &        1.5                 &  0.0                \\ 
       Resolution          & 0.0                  & 0.2                      &   0.0         &        0.2                 &  0.0                \\ 
       Mass scale          & 0.1                  & 0.0                      &   0.1         &        0.0                 &  0.1                \\ 
       Interference        & 0.1                  & 0.3                      &   0.1         &        0.1                 &  0.1                \\ \hline
       Total               & 0.6                  & 1.1                      &   0.2         &        1.5                 &  0.1                \\ \hline \hline
    \end{tabular}    
    \label{summary_sys_lamd0}
  \end{table*}
\end{center}

\begin{center}
  \begin{table*}[htbp]
    \caption{Summary of the masses, widths and significances of the $\Xi_{c}^{\ast}$ baryons measured in the $\Lambda D$ modes.
             The first error is statistical and the second is systematic.
             We set a 90$\%$ confidence level upper limit for the width of $\Xi_c(3080)^{+}$.}
     \begin{tabular}{c|ccc} \hline \hline
       Resonance           &  Mass (MeV/$c^{2}$)     & Width (MeV)          & Significance ($\sigma$) \\ \hline
       $\Xi_c(3055)^{0}$   &  $3059.0\pm0.5\pm0.6$   & $6.4\pm2.1\pm1.1$    & 8.6  \\
       $\Xi_c(3055)^{+}$   &  $3055.8\pm0.4\pm0.2$   & $7.0\pm1.2\pm1.5$    & 11.7 \\
       $\Xi_c(3080)^{+}$   &  $3079.6\pm0.4\pm0.1$   &  $<$ 6.3             & 4.8  \\ \hline \hline
    \end{tabular}    
    \label{summary_mass_lambdad}
  \end{table*}
\end{center}

\section{Combined analysis with the $\Sigma_{c}^{++} K^{-}$ and $\Sigma_{c}^{\ast ++} K^{-}$ decay modes}\label{section_combine}
\begin{figure*}[htbp]
  \begin{center}
    \includegraphics[scale=0.3]{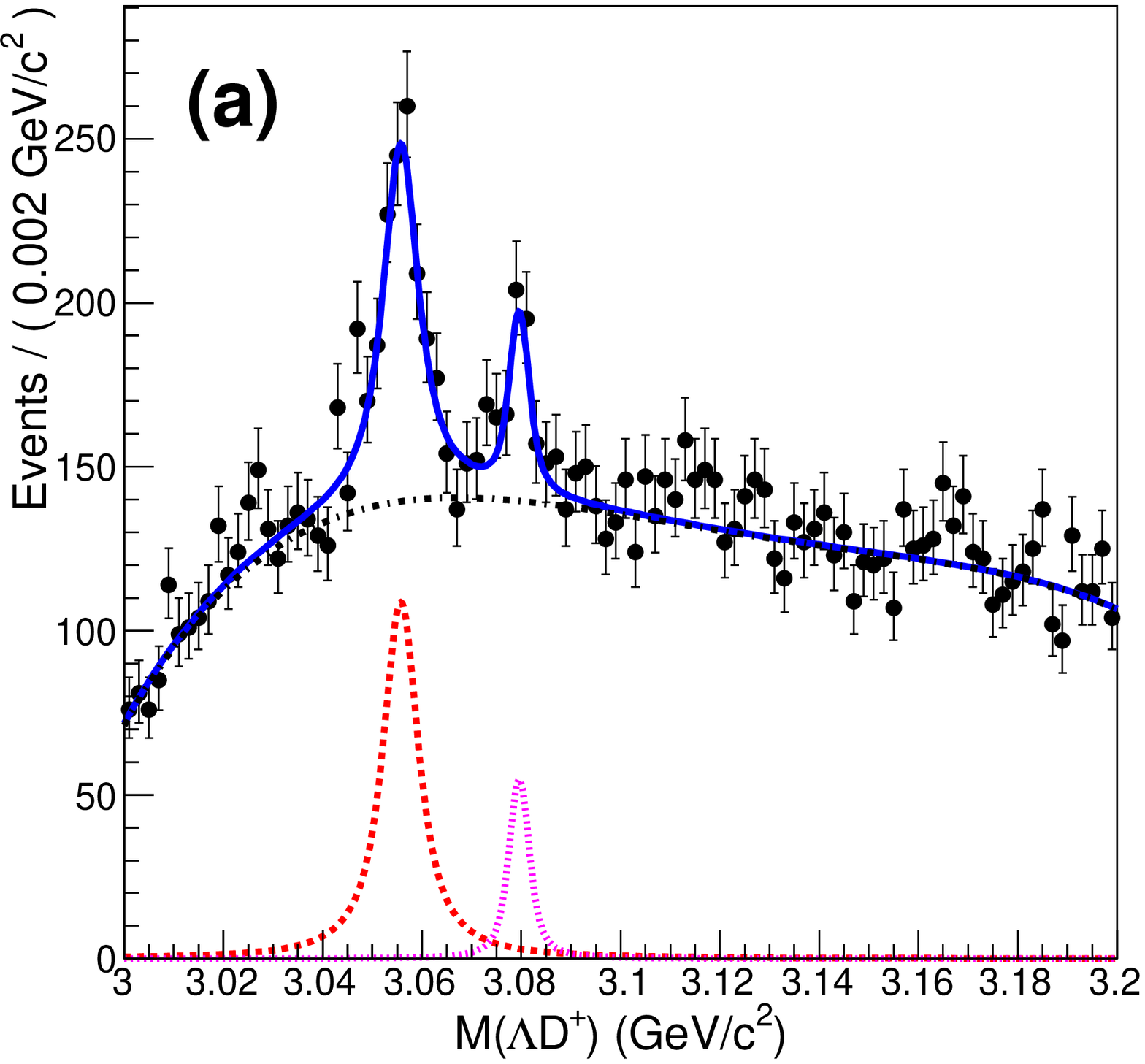}
    \includegraphics[scale=0.3]{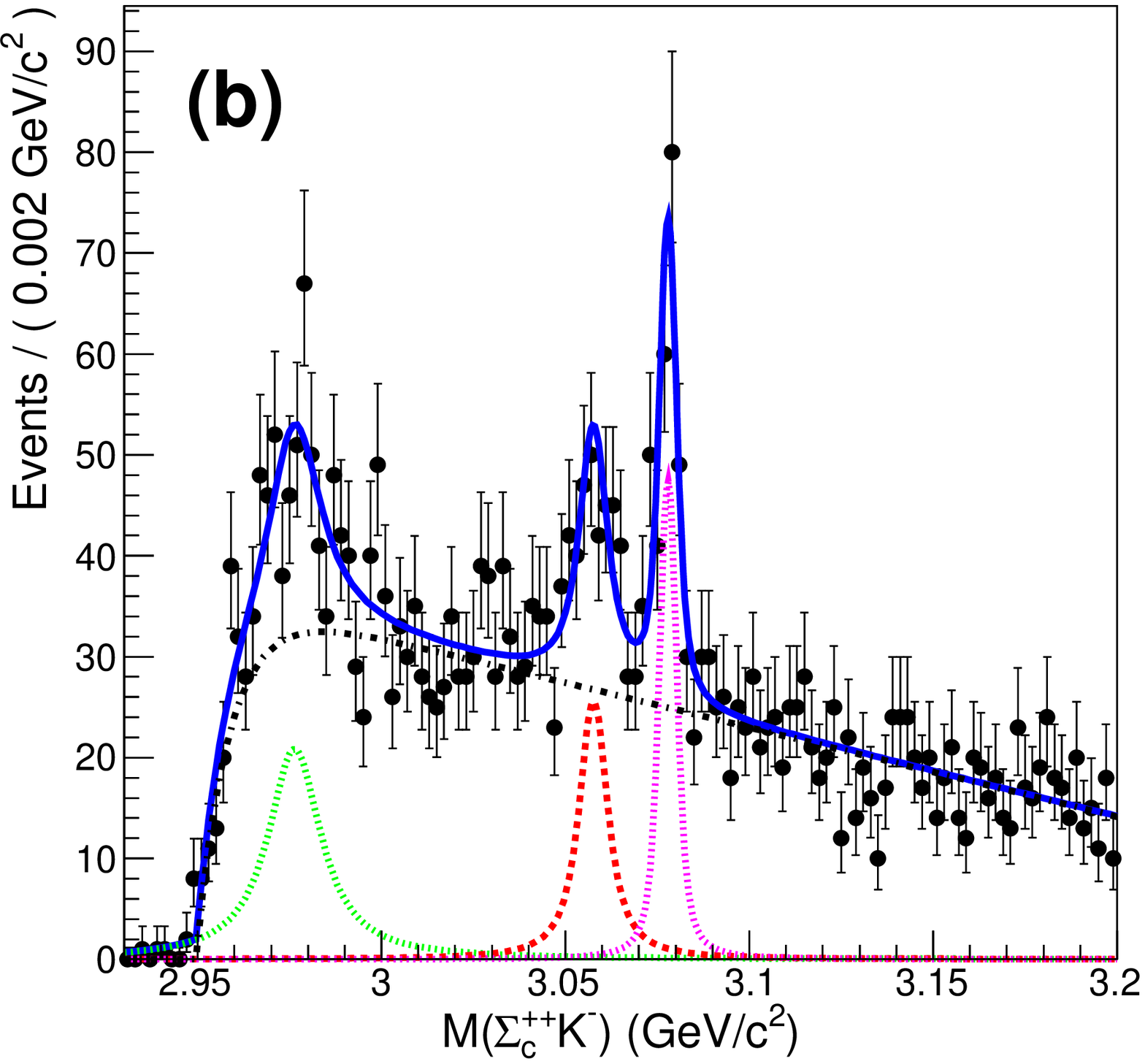}
    \includegraphics[scale=0.3]{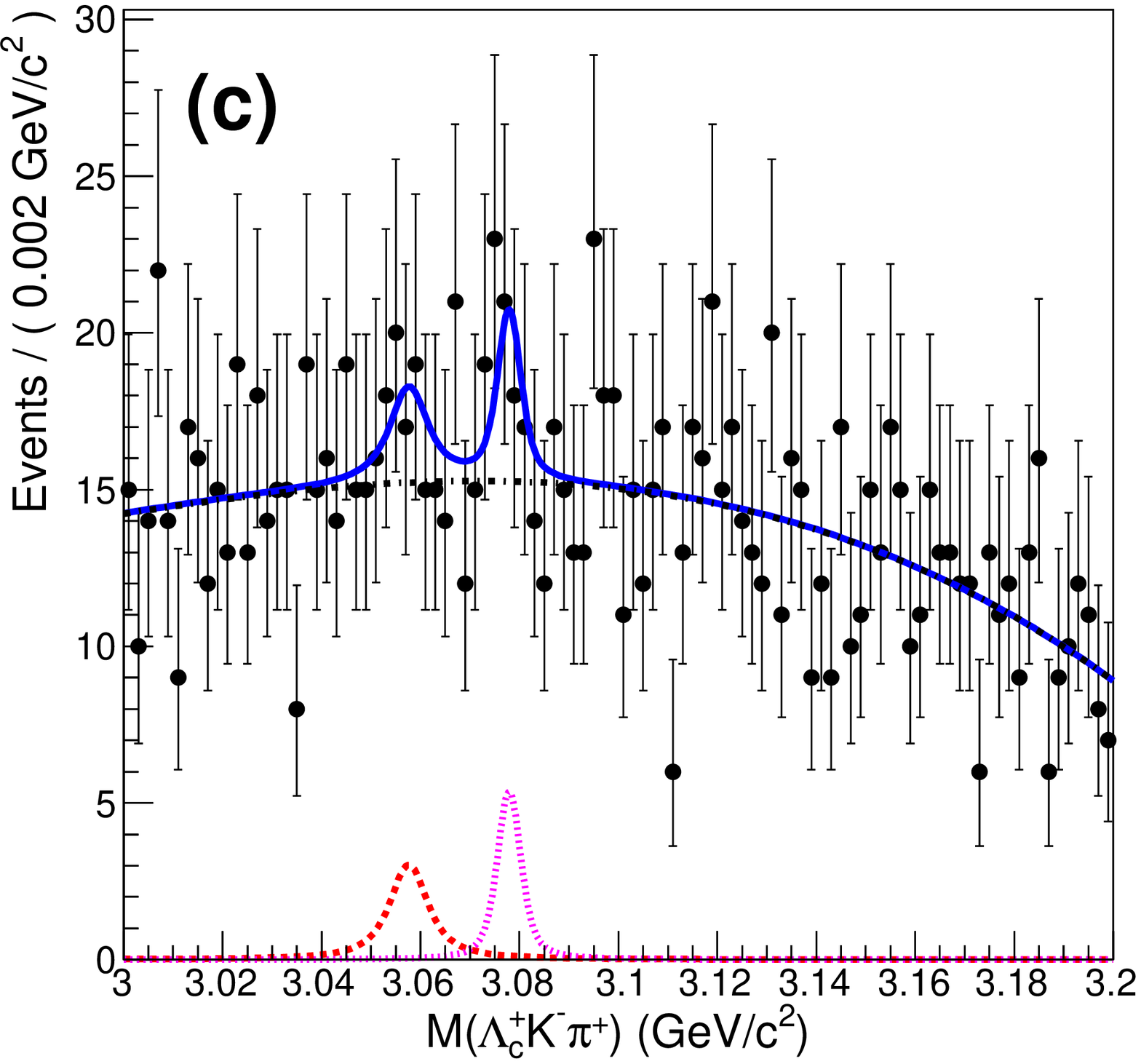}
    \includegraphics[scale=0.3]{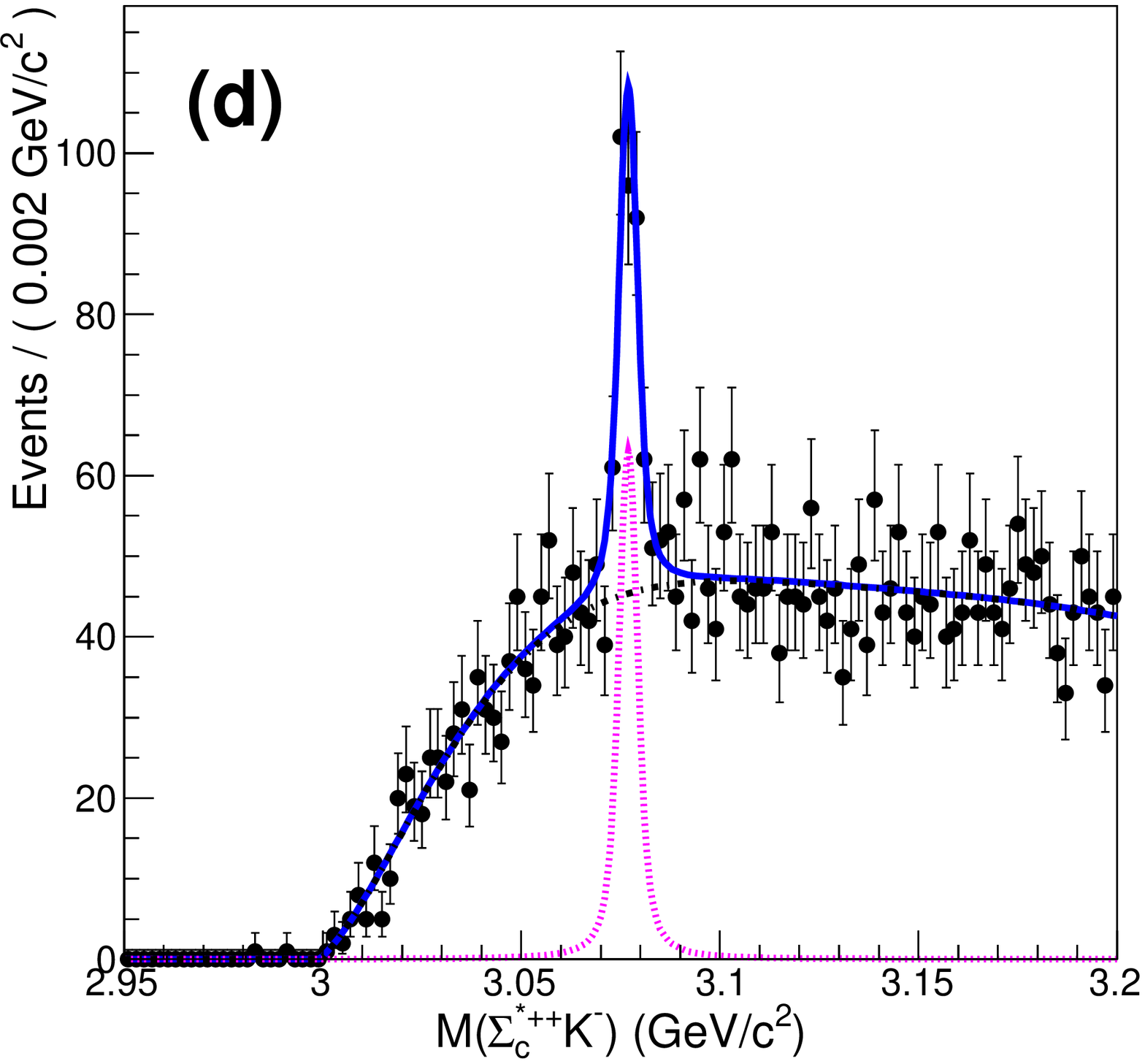}
    \caption{The simultaneous fit results. Points with error bars are data.
             The blue solid lines show the fit result. The red dashed, magenta dotted, green dotted, and black dash-dotted lines
	     show the contributions from the $\Xi_{c}(3055)^{+}$, $\Xi_{c}(3080)^{+}$, $\Xi_{c}(2980)^{+}$, and background, respectively.
             (a) $M(\Lambda D^{+}$), (b) $M(\Sigma_{c}^{++}K^{-}$), (c) $M(\Lambda_{c}^{+}K^{-} \pi^{+}$)
             for the $\Sigma_{c}^{++}$ sideband region, and (d) $M(\Sigma_{c}^{\ast ++}K^{-})$.}
    \label{simultaneousfit}
  \end{center}
\end{figure*}

We measure the ratio of branching fractions,
${{\cal B}(\Xi_{c}^{\ast +} \to \Lambda D^{+})}/{{\cal B}(\Xi_{c}^{\ast +} \to \Sigma_{c}^{++}K^{-})} \equiv R_{{\cal B} (\Lambda D)}$,
using the following equations:
\begin{eqnarray}
\lefteqn{R_{{\cal B} (\Lambda D)}=R_{{\rm yield} (\Lambda D)}\times({\cal B} \times \epsilon)_{\Sigma_{c}K}/({\cal B} \times \epsilon)_{\Lambda D^{+}}}, \\
\lefteqn{({\cal B} \times \epsilon)_{\Lambda D^{+}}={\cal B}(D^{+} \to K^{-}\pi^{+}\pi^{+})  } \nonumber \\
\lefteqn{\:\:\:\:\:\:\:\:\:\:\:\:\:\:\:\:\:\:\:\: \:\:\:\:\:\:\: \times {\cal B}(\Lambda \to p \pi^{-}) \times \epsilon(\Lambda D^{+}) }  \\
\lefteqn{({\cal B} \times \epsilon)_{\Sigma_{c}K}={\cal B}(\Lambda_{c}^{+}\to pK^{-}\pi^{+}) \times[ \epsilon_{p K^{-}\pi^{+}} + R_{pK^{0}_{S}} } \nonumber \\ 
\lefteqn{\:\:\:\:\:\:\:\:\:\:\:\:\:\:\:\:\: \:\:\:\:\: \:\:\:\:\: \times {\cal B}(K^{0}_{S} \to \pi^{+}\pi^{-}) \times \epsilon_{p K^{0}_{S}} ], } 
\end{eqnarray}
where $\epsilon(\Lambda D^{+})$ is the reconstruction efficiency for the $\Lambda D^{+}$ mode, $\epsilon_{i}$ is the 
reconstruction efficiency for the $\Sigma_{c}^{++}K^{-}$ mode with the $i^{\rm th}$ sub-decay of the $\Lambda_{c}^{+}$,
$R_{pK^{0}_{S}}$ is the ratio of branching fraction ${\cal B}( \Lambda_{c}^{+} \to p K^{0}_{S})/{\cal B}( \Lambda_{c}^{+} \to p K^{-} \pi^{+})$
and $R_{{\rm yield} (\Lambda D)}$ is the ratio of the yields of $\Xi_{c}^{\ast}$ baryons in the $\Lambda D^{+}$ and the $\Sigma_{c}^{++}K^{-}$ modes.
For ${\cal B}(\Lambda_{c}^{+}\to pK^{-}\pi^{+})$, we use the latest Belle measurement \cite{Zupanc:2013iki}.
Other branching fraction values are taken from Ref.~\cite{PDG}. 
We also measure the ratio of branching fractions,
${{\cal B}(\Xi_{c}(3080)^{+} \to \Sigma_c^{\ast ++} K^{-})}/{{\cal B}(\Xi_{c}(3080)^{+} \to \Sigma_{c}^{++}K^{-})}$ $\equiv R_{{\cal B}\Sigma_{c}^{\ast} K}$
using the equation
\begin{eqnarray}
\lefteqn{R_{{\cal B} (\Sigma_{c}^{\ast} K)}=R_{{\rm yield} (\Sigma_{c}^{\ast} K)}\times({\cal B} \times \epsilon)_{\Sigma_{c}K}/({\cal B} \times \epsilon)_{\Sigma_{c}^{\ast} K}},
\end{eqnarray}
where $R_{{\rm yield} (\Sigma_{c}^{\ast} K)}$ is the ratio of yields of $\Xi_{c}(3080)^{+}$ in the $\Sigma_c^{\ast ++} K^{-}$ decay mode and $\Sigma_{c}^{++}K^{-}$ decay modes.
$({\cal B} \times \epsilon)_{\Sigma_{c}^{\ast} K}$ shares the form of Eq. (3) for
$({\cal B} \times \epsilon)_{\Sigma_{c}K}$ after replacing the reconstruction efficiency
for $\Sigma_{c}^{\ast ++}K^{-}$ with that for $\Sigma_{c}^{++}K^{-}$.
The data set used for the $\Sigma_{c}^{++}K^{-}$ and $\Sigma_{c}^{\ast ++}K^{-}$ decay modes is the same as that for the $\Lambda D^{+}$ mode.
Event selections are the same as those in Ref.~\cite{Kato:2013ynr}.
A $\Sigma_{c}^{++}$ or $\Sigma_{c}^{\ast ++}$ candidate is reconstructed via its decay into $\Lambda_{c}^{+} \pi^{+}$; the $\Lambda_{c}^{+}$ candidate here is reconstructed via its decay into
$p K^{-} \pi^{+}$ and $p K^{0}_{S}$. Note that the requirement $x_{p}>0.7$ is the same as that for the $\Lambda D^{+}$ mode
and so it is possible to directly compare the three decay modes.
To obtain $R_{\rm yield}$ and to measure the width of the $\Xi_{c}(3055)^{+}$ and $\Xi_{c}(3080)^{+}$ with greater
accuracy than is possible using a single decay mode, we perform a simultaneous UML fit with the widths of the  
$\Xi_{c}^{\ast}$ states constrained to be the same among the three decay modes, as discussed in the previous section.
The masses are not constrained because we find inconsistency for the mass of the $\Xi_{c}(3080)^{+}$ among the three decay modes.
We also fit the mass distribution of the $\Sigma_{c}^{++}$ sideband region, defined as
$|M(\Lambda_c^{+}\pi^{+})-(m_{\Sigma_c^{++}} \pm 15 $ MeV/${\it c}$$^{2}$$)| \: <$ 5 MeV/${\it c}$$^{2}$,
where $m_{\Sigma_c^{++}}$ is the $\Sigma_c^{++}$ mass, to subtract the contribution from non-resonant $\Lambda_{c}^{+} K^{-} \pi^{+}$ decays
in the signal region.
We subtract half of the yield found in the sideband regions because the mass range of the sideband region is double
the width of the $\Sigma_{c}^{++}$ signal region. 
It is difficult to define the $\Sigma_{c}^{\ast ++}$ sideband regions because the maximum
mass that is possible for combinations to contribute to the $\Xi_{c}(3080)^{+}$ is only slightly higher 
than the $\Sigma_{c}^{\ast ++}$ mass, and a low mass sideband would overlap with the $\Sigma_{c}^{++}$ region.
Thus, we estimate the contribution under the $\Sigma_{c}^{\ast ++}$ by scaling the yield in the 
$\Sigma_{c}^{++}$ sideband regions by 2.9, a factor estimated using signal MC.
We assume no interference between $\Sigma_{c}^{++} K^{-}$ or $\Sigma_{c}^{\ast ++} K^{-}$ with non-resonant $\Lambda_{c}^{+}K^{-}\pi^{+}$.
The PDFs and fit region for the $\Lambda D^{+}$ are the same as those described in Section \ref{section_signal}.
The fit conditions for the $\Sigma_{c}^{++}K^{-}$ and $\Sigma_{c}^{\ast ++}K^{-}$ modes are the same as in Ref.~\cite{Kato:2013ynr}.
For the fit to the events from the $\Sigma_{c}^{++}$ sideband region, we use the $3.0-3.2$ ${\rm GeV}/c^{2}$ $\Sigma_{c}^{++}K^{-}$ mass range.
The $\Xi_{c}^{\ast}$ contributions are represented by a Gaussian-convolved Breit-Wigner
with the same mass resolution of the $\Xi_{c}^{\ast}$ states as that used for the $\Sigma_{c}^{++}$ signal region.
The combinatorial background is represented by a second-order Chebyshev polynomial.
Figure \ref{simultaneousfit} shows the results of the simultaneous fit.

The following systematic uncertainties are taken into account for the combined analysis
for the measurements of the ratios of branching fractions and width.
The systematic uncertainty due to the pion- and kaon-identification efficiency
is estimated from the ratio of the yields of the
$D^{\ast +} \to D^{0}\pi^{+}$, $D^{0}\to K^{-}\pi^{+}$ with and without 
the pion- and kaon-identification requirements for data and MC.
The difference of the ratio between data and MC is used to correct the efficiency
and the statistical error of this correction is treated as the systematic uncertainty.
We conservatively assume no correlation in the systematic uncertainty for pion and kaon identification between
$\Lambda D^{+}$ and $\Sigma_{c}^{++}K^{-}$ decay modes as the momentum ranges for these
decay modes are distinct; the systematic uncertainty for 
$\Sigma_{c}^{++}K^{-}$ and $\Sigma_{c}^{\ast ++}K^{-}$ cancel.
The systematic uncertainty due to the efficiency of proton identification 
is determined using the ratio of the yields of the 
$\Lambda \to p \pi^{-}$ with and without the proton identification requirement.
The difference of the ratio between data and MC is used to correct the efficiency 
and the statistical uncertainty of this correction is regarded as the systematic uncertainty.
The systematic uncertainty due to the reconstruction efficiency of the $\Lambda$ is determined using the yield ratio of 
$B \to \Lambda \bar{\Lambda} K^{+}$ with and without the $\Lambda$ selection cut as a function of momenta of $\Lambda$.
By taking the weighted average of the momentum, it is estimated to be 3$\%$.
The uncertainties of the branching fractions \cite{PDG,Zupanc:2013iki} are included as systematic uncertainties.
The stability of the background shape is checked by changing the fit region and background PDF.
The maximum deviation from the nominal fit among the various changes is regarded as the systematic uncertainty.
To assess the uncertainty due to $\sigma_{\rm res}$, $r_{\sigma}$ is evaluated as $\sigma_{D}^{MC}/ \sigma_{D}^{data}=1.15$
for the $\Lambda D^{+}$ mode and $\sigma_{\Lambda_{c}^{+}}^{MC}/\sigma_{\Lambda_{c}^{+}}^{data}=1.08$
for the $\Sigma_{c}^{++}K^{-}$; we perform a fit with $\sigma_{\rm res}$ scaled by a factor of $r_{\sigma}$
and use the difference of the result from the nominal fit as the systematic uncertainty.
To check the uncertainty due to a possible mis-calibration of momentum and energy measurements, 
we evaluate the difference between the reconstructed and nominal $D^{+}$ and $\Lambda_{c}^{+}$ masses for both data and MC.
In data, the reconstructed $D^{+}$ mass differs from the world average \cite{PDG} by 0.1 MeV/$c^{2}$
whereas, in the MC, the $D^{+}$ mass differs by 0.2 MeV/$c^2$. No deviation is observed for $\Lambda_{c}^{+}$
for both data and MC. In the signal MC, the difference of the input and output $\Xi_{c}^{\ast}$ masses
in the $\Lambda D^{+}$ mode is 0.1 MeV/$c^{2}$, which is smaller than the deviation observed in the $D^{+}$ mass because of the
mass-constrained fit. We conservatively assign the systematic uncertainty of 0.1 MeV/$c^{2}$ on the mass measurement.
Table \ref{summary_sys_rel} summarizes the systematic uncertainties. 
Table \ref{summary_simfit} summarizes the measurement of yields and widths of the $\Xi_{c}(3055)^{+}$ and $\Xi_{c}(3080)^{+}$ and 
Table \ref{summary_branch_ratio} summarizes the values related to the ratio of branching fractions measurements.

%\begin{center}
%  \begin{table*}[htbp]
%    \caption{Summary of the systematic uncertainties for the width (MeV) and ratio of branching fraction ratios ($\%$) measurements from the combined analysis.}
%     \begin{tabular}{c|ccccc} \hline \hline
%       Source                                           & $\Gamma_{\Xi_c(3055)^{+}}$ & $R_{{\cal B} (\Lambda D)}$ ($\Xi_c(3055)^{+}$)  & $\Gamma_{\Xi_c(3080)^{+}}$ & $R_{{\cal B} (\Lambda D)}$ ($\Xi_c(3080)^{+}$) & $R_{{\cal B} (\Sigma_c^{\ast} K)}$\\ \hline
%       Particle identification ($\Lambda D^{+}$ mode)               &-      &3.3   &-    &3.3   &  -   \\
%       Particle identification ($\Sigma_{c}^{++}K^{-}$ mode)        &-      &2.0   &-    &2.0   &  -   \\
%       Branching fractions                                          &-      &5.7   &-    &5.7   &  -   \\
%       Background shape                                             &1.5    &13.1  &0.4  &9.7   &  1.0 \\
%       Resolution                                                   &0.2    &2.1   &0.2  &1.6   &  0.5 \\
%       Mass scale                                                   &0.0    &0.0   &0.0  &0.0   &  0.0 \\ \hline
%       Total                                                        &1.5    &14.9  &0.5  &12.0  &  1.1 \\ \hline \hline 
%    \end{tabular}    
%    \label{summary_sys_rel}
%  \end{table*}
%\end{center}

\begin{center}
  \begin{table*}[htbp]
    \caption{Summary of the systematic uncertainties for the width (MeV) and ratio of branching fraction ratios ($\%$) measurements from the combined analysis.}
     \begin{tabular}{c|ccccc} \hline \hline
       Source                                  & $\Gamma_{\Xi_c(3055)^{+}}$ & $R_{{\cal B} (\Lambda D)}$ for $\Xi_c(3055)^{+}$  & $\Gamma_{\Xi_c(3080)^{+}}$ & $R_{{\cal B} (\Lambda D)}$ for $\Xi_c(3080)^{+}$ & $R_{{\cal B} (\Sigma_c^{\ast} K)}$\\ \hline
       $\pi K p$ identification                &-      &1.4   &-    &1.4   &  -   \\
       $\Lambda$ identification                &-      &3.0   &-    &3.0   &  -   \\
       Branching fractions                     &-      &5.7   &-    &5.7   &  -   \\
       Background shape                        &1.5    &13.1  &0.4  &9.7   &  1.0 \\
       Resolution                              &0.2    &2.1   &0.2  &1.6   &  0.5 \\
       Mass scale                              &0.0    &0.0   &0.0  &0.0   &  0.0 \\ \hline
       Total                                   &1.5    &14.9  &0.4  &12.0  &  1.1 \\ \hline \hline 
    \end{tabular}    
    \label{summary_sys_rel}
  \end{table*}
\end{center}

\begin{center}
  \begin{table*}[htbp]
     \caption{Summary of results from the simultaneous fits to the $\Lambda D^{+}$ and $\Sigma_{c}^{++}K^{-}$ modes.}
     \begin{tabular}{c|ccccc} \hline \hline
       Resonance             & Width (MeV)      & Yield for $\Lambda D^{+}$ & Yield for $\Sigma_{c}^{++}K^{-}$ & Yield for sideband & Yield for $\Sigma_{c}^{\ast ++}$ \\ \hline
       $\Xi_{c}(3055)^{+}$   & $7.8\pm1.2\pm1.5$ &  $721\pm90$               & $173\pm30$                    & $21\pm18$        &   -\\ 
       $\Xi_{c}(3080)^{+}$   & $3.0\pm0.7\pm0.4$ &  $186\pm40$               & $176\pm23$                    & $20\pm12$        &   $234\pm30$\\  \hline \hline
    \end{tabular}    
    \label{summary_simfit}
  \end{table*}
\end{center}

\begin{center}
  \begin{table*}[htbp]
    \caption{Summary of the values related to the measurements of the ratio of branching fractions.
             The branching fraction values are taken from Ref.~\cite{PDG,Zupanc:2013iki}.
	     For the ratios of branching fractions, the first error is statistical and second is systematic.}
     \begin{tabular}{c|c} \hline \hline
       Variable                                                                             &  Value                 \\ \\ \hline
       ${\cal B}(D^{+} \to K^{-}\pi^{+}\pi^{+})$                                               &  $0.0913\pm0.0019$     \\ 
       ${\cal B}(\Lambda \to p \pi^{-})$                                                     &  $0.639\pm0.005$       \\ 
       ${\cal B}(\Lambda_{c}^{+}\to pK^{-}\pi^{+})$                                            &  $0.0684\pm0.036$      \\ 
       ${\cal B}(K^{0}_{S}\to\pi^{+}\pi^{-})$                                                  &  $0.6920\pm0.0005$     \\
       ${\cal B}(\Lambda_{c}^{+}\to p K^{0}_{S})/{\cal B}(\Lambda_{c}^{+}\to pK^{-}\pi^{+})$      &  $0.24\pm0.02$         \\
       $\epsilon(\Lambda D^{+})$                                                             &  $0.1771$        \\
       $\epsilon_{p K^{-}\pi^{+}}$ ($\Sigma_{c}^{++}K^{-}$)                                        &  $0.149$               \\ 
       $\epsilon_{p K^{0}_{S}}$    ($\Sigma_{c}^{++}K^{-}$)                                        &  $0.155$               \\
       $\epsilon_{p K^{-}\pi^{+}}$ ($\Sigma_{c}^{\ast ++}K^{-}$)                                     &  $0.146$               \\ 
       $\epsilon_{p K^{0}_{S}}$    ($\Sigma_{c}^{\ast ++}K^{-}$)                                    &  $0.153$               \\
       $({\cal B} \times \epsilon)_{\Lambda D^{+}}$                                             &  $0.0103$              \\ 
       $({\cal B} \times \epsilon)_{\Sigma_{c}K}$                                               &  $0.0119$              \\
       $({\cal B} \times \epsilon)_{\Sigma_{c}^{\ast}K}$                                           &  $0.0117$              \\
       $R_{{\rm yield} (\Lambda D)}$ for $\Xi_c(3055)^{+}$                                           &  $4.41\pm0.87$         \\ 
       $R_{{\rm yield} (\Lambda D)}$ for $\Xi_c(3080)^{+}$                                           &  $1.12\pm0.26$         \\ 
       $R_{{\rm yield} (\Sigma_{c}^{\ast} K)}$                                                          &  $1.05\pm0.27$         \\ 
       $R_{{\cal B} (\Lambda D)}$ for $\Xi_c(3055)^{+}$                                            &  $5.09\pm1.01\pm0.76$  \\ 
       $R_{{\cal B} (\Lambda D)}$ for $\Xi_c(3080)^{+}$                                            &  $1.29\pm0.30\pm0.15$  \\ 
       $R_{{\cal B} (\Sigma_{c}^{\ast}K)}$                                                           &  $1.07\pm0.27\pm0.01$  \\ \hline \hline
    \end{tabular}
    \label{summary_branch_ratio}
  \end{table*}
\end{center}

\section{Summary and conclusions}\label{section_conclusion}
We present studies of $\Xi_{c}^{\ast}$ baryons decaying into the $\Lambda D^{+}$ and $\Lambda D^{0}$ final states.
We report the first observation of the $\Xi_{c}(3055)^{0}$ in the $\Lambda D^{0}$ mode with a significance of $8.6 \sigma$.
The mass and width of the $\Xi_{c}(3055)^{0}$ are measured to be (3059.0$\pm$0.5$\pm$0.6) MeV/$c^{2}$ and (6.4$\pm$2.1$\pm$1.1) MeV,
respectively. We report the first observation of the $\Xi_{c}(3055)^{+}$ decay and evidence for the $\Xi_{c}(3080)^{+}$ 
in the $\Lambda D^{+}$ final state. 
The mass and width of the $\Xi_{c}(3055)^{+}$ obtained from the $\Lambda D$ final states only are
($3055.8\pm0.4\pm0.2$) MeV/$c^{2}$ and $(7.0\pm1.2\pm1.5)$ MeV, respectively, and those for $\Xi_c(3080)^{+}$ are 
($3079.6\pm0.4\pm0.1$) MeV/$c^{2}$ and $<$ 6.3 MeV, respectively.
The measured values for $\Xi_{c}(3055)^{+}$ are more accurate than the world average thanks to the high statistics in this decay mode.

We perform a combined analysis of these particles by comparing their decays into $\Lambda D^{+}$ with those into $\Sigma_{c}^{++}K^{-}$ and $\Sigma_{c}^{\ast ++}K^{-}$.
We measure the ratios of branching fractions ${\cal B}(\Xi_{c}(3055)^{+} \to \Lambda D^{+})/{\cal B}(\Xi_{c}(3055)^{+} \to \Sigma_{c}^{++}K^{-})=5.09\pm1.01\pm0.76$,
${\cal B}(\Xi_{c}(3080)^{+} \to \Lambda D^{+})/{\cal B}(\Xi_{c}(3080)^{+} \to \Sigma_{c}^{++}K^{-})=1.29\pm0.30\pm0.15$,
and ${\cal B}(\Xi_{c}(3080)^{+} \to \Sigma_{c}^{\ast ++}K^{-} )/{\cal B}(\Xi_{c}(3080)^{+} \to \Sigma_{c}^{++}K^{-})=1.07\pm0.27\pm0.01$.
The width of the $\Xi_{c}(3055)^{+}$ is $(7.8\pm1.2\pm1.5)$ MeV and that of the $\Xi_{c}(3080)^{+}$ is $(3.0\pm0.7\pm0.4)$ MeV.
We take the weighted average of the measurements in the different decay modes to find the masses of the $\Xi_{c}(3055)^{+}$ and $\Xi_{c}(3080)^{+}$
to be $(3055.9\pm0.4)$ MeV/$c^{2}$ and $(3077.9\pm0.9)$ MeV/$c^{2}$, respectively, 
where the uncertainties are scaled by $\sqrt{\chi^{2}/(N-1)}$ to account for small inconsistencies in the $N$ individual measurements.
The uncertainties on the masses incorporate the statistical and systematic values.
The masses and widths of $\Xi_{c}(3055)^{+}$ and $\Xi_{c }(3080)^{+}$, after combining other decay modes,
supersede our previous measurements \cite{Kato:2013ynr}.

Our measurements provide information on the nature of these baryons.
For instance, the chiral quark model has been used to identify the $\Xi_c(3055)$ as the D-wave excitation in the 
N=2 shell, and predicts ${\cal B} (\Xi_{c}(3055) \to \Sigma_c \bar{K})$:${\cal B} (\Xi_{c}(3055) \to \Lambda D)$ to be 2.3:0.1 or 5.6:0.0,
depending on the possible excitation modes \cite{Liu:2012sj}.
It further identifies the $\Xi_c(3080)$ as an S-wave excitation mode of the $\Xi_{c}$ in N=2 shell 
and predicts that its decay into $\Lambda D$ is forbidden. 
Both of these predictions are in contradiction with our measurements.
Further experimental and theoretical work is needed to understand these baryons.

\acknowledgments

We thank the KEKB group for the excellent operation of the
accelerator; the KEK cryogenics group for the efficient
operation of the solenoid; and the KEK computer group,
the National Institute of Informatics, and the 
PNNL/EMSL computing group for valuable computing
and SINET4 network support.  We acknowledge support from
the Ministry of Education, Culture, Sports, Science, and
Technology (MEXT) of Japan, the Japan Society for the 
Promotion of Science (JSPS), and the Tau-Lepton Physics 
Research Center of Nagoya University; 
the Australian Research Council;
Austrian Science Fund under Grant No.~P 22742-N16 and P 26794-N20;
the National Natural Science Foundation of China under Contracts 
No.~10575109, No.~10775142, No.~10875115, No.~11175187, No.~11475187
and No.~11575017;
the Chinese Academy of Science Center for Excellence in Particle Physics; 
the Ministry of Education, Youth and Sports of the Czech
Republic under Contract No.~LG14034;
the Carl Zeiss Foundation, the Deutsche Forschungsgemeinschaft, the
Excellence Cluster Universe, and the VolkswagenStiftung;
the Department of Science and Technology of India; 
the Istituto Nazionale di Fisica Nucleare of Italy; 
the WCU program of the Ministry of Education, National Research Foundation (NRF) 
of Korea Grants No.~2011-0029457,  No.~2012-0008143,  
No.~2012R1A1A2008330, No.~2013R1A1A3007772, No.~2014R1A2A2A01005286, 
No.~2014R1A2A2A01002734, No.~2015R1A2A2A01003280 , No. 2015H1A2A1033649;
the Basic Research Lab program under NRF Grant No.~KRF-2011-0020333,
Center for Korean J-PARC Users, No.~NRF-2013K1A3A7A06056592; 
the Brain Korea 21-Plus program and Radiation Science Research Institute;
the Polish Ministry of Science and Higher Education and 
the National Science Center;
the Ministry of Education and Science of the Russian Federation and
the Russian Foundation for Basic Research;
the Slovenian Research Agency;
Ikerbasque, Basque Foundation for Science and
the Euskal Herriko Unibertsitatea (UPV/EHU) under program UFI 11/55 (Spain);
the Swiss National Science Foundation; 
the Ministry of Education and the Ministry of Science and Technology of Taiwan;
and the U.S.\ Department of Energy and the National Science Foundation.
This work is supported by a Grant-in-Aid for Scientific Research (S) ''Probing New Physics with Tau-Lepton'' (No.26220706),
Grant-in-Aid for Scientific Research on Innovative Areas ''Elucidation of New Hadrons with a Variety of Flavors'',
Grant-in-Aid from MEXT for Science Research in a Priority Area (``New Development of Flavor Physics'')
and from JSPS for Creative Scientific Research (``Evolution of Tau-lepton Physics'').

\end{document}